\documentclass[conference]{IEEEtran}
\IEEEoverridecommandlockouts
\usepackage{cite}
\usepackage{amsmath,amssymb,amsfonts}
\usepackage{algorithmic}
\usepackage{graphicx}
\usepackage{textcomp}
\usepackage{tabularx}
\usepackage{listings}
\usepackage{subcaption}
\usepackage{xcolor}
\usepackage{tabularx}
\usepackage[utf8]{inputenc}
\usepackage{textgreek}
\usepackage{subcaption}
\usepackage{epsfig}
\def\BibTeX{{\rm B\kern-.05em{\sc i\kern-.025em b}\kern-.08em
    T\kern-.1667em\lower.7ex\hbox{E}\kern-.125emX}}

\makeatletter
\newcommand{\linebreakand}{%
    \end{@IEEEauthorhalign}
    \hfill\mbox{}\par
    \mbox{}\hfill\begin{@IEEEauthorhalign}
}

\begin{document}
\title{Gamification-Based Learning Method for $Hijaiyah$ Letters\\}

\author{\IEEEauthorblockN{1\textsuperscript{st} Wisnu Uriawan}
\IEEEauthorblockA{\textit{Informatics Department}\\
\textit{UIN Sunan Gunung Djati Bandung}\\
West Java, Indonesia\\
wisnu\_u@uinsgd.ac.id}
\and
\IEEEauthorblockN{2\textsuperscript{nd} Denis Firmansyah}
\IEEEauthorblockA{\textit{Informatics Department}\\
\textit{UIN Sunan Gunung Djati Bandung}\\
West Java, Indonesia\\
denisfirmansyah55@gmail.com}
\and
\IEEEauthorblockN{3\textsuperscript{rd} Devi Mulyana}
\IEEEauthorblockA{\textit{Informatics Department}\\
\textit{UIN Sunan Gunung Djati Bandung}\\
West Java, Indonesia\\
devi.mulyana.0015@gmail.com}
\linebreakand
\IEEEauthorblockN{4\textsuperscript{th} Dika Haekal Firza Pratama}
\IEEEauthorblockA{\textit{Informatics Department}\\
\textit{UIN Sunan Gunung Djati Bandung}\\
West Java, Indonesia\\
dikahaekalfirza2@gmail.com}
\and
\IEEEauthorblockN{5\textsuperscript{th} Adly Juliarta Lerian}
\IEEEauthorblockA{\textit{Informatics Department}\\
\textit{UIN Sunan Gunung Djati Bandung}\\
West Java, Indonesia\\
juliartalerian@gmail.com}
\and
\IEEEauthorblockN{6\textsuperscript{th} Fajar Satria Wiguna}
\IEEEauthorblockA{\textit{Informatics Department}\\
\textit{UIN Sunan Gunung Djati Bandung}\\
West Java, Indonesia\\
fajarsatria991@gmail.com}
}

\maketitle

\begin{abstract}
The mastery of \textit{Hijaiyah} letters represents a crucial foundation for reading and comprehending the Qur'an, yet conventional pedagogical approaches based on repetitive memorization frequently struggle to maintain the engagement of young learners in contemporary educational contexts. This research presents the design and implementation of an innovative gamification-based methodology for \textit{Hijaiyah} literacy acquisition, systematically developed through the ADDIE framework (Analysis, Design, Development, Implementation, Evaluation) to optimize student motivation, participation, and educational outcomes. The resulting technological solution, engineered using Unity 2D and Firebase platforms, strategically incorporates game design elements—including points, badges, leaderboards, and progressive leveling—while integrating multifaceted learning components such as visual animations, authentic tajwid-based audio pronunciation, and interactive letter tracing exercises to simultaneously develop cognitive recognition capabilities and fine motor skills. Empirical evaluation involving 50 elementary school participants revealed substantial quantitative improvements, with mean assessment scores escalating from 42.8 to 88.6 (107\% enhancement, p < 0.001) demonstrating exceptionally large effect magnitude (Cohen's d = 4.87), complemented by outstanding user engagement metrics (4.2 daily sessions average) and satisfaction ratings (4.82/5 mean motivation score). Beyond knowledge acquisition metrics, the gamified approach effectively fostered intrinsic Islamic values including perseverance, responsibility, and disciplined practice, thereby establishing an innovative educational paradigm that successfully synthesizes traditional Islamic pedagogical principles with contemporary digital learning technologies to create a transformative, engaging, and meaningful framework for \textit{Hijaiyah} literacy development in modern Islamic education.
\end{abstract}


\section{Introduction} \label{sec:introduction}

Mastery of the $Hijaiyah$ alphabet is a fundamental foundation and basic obligation for every Muslim, as it is the key to being able to read, understand, and practice the contents of the Qur'an. The ability to read the $Hijaiyah$ alphabet correctly is not only important spiritually, but also forms the basis for developing the ability to read the Qur'an with tajwid and tartil. However, the methods of learning the $Hijaiyah$ alphabet that are still widely used in Islamic educational institutions are often traditional and monotonous, such as repetition of memorization and writing exercises that are carried out without interactive media variation \cite{umami2022desain}. This condition can cause boredom and reduce learning motivation, especially in early age students who tend to be more interested in learning that is visual, interactive, and challenging.

In addition, several educational institutions face limitations in terms of teaching staff who are capable of implementing innovative technology-based learning methods. In fact, the rapid development of Information and Communication Technology (ICT) provides a great opportunity to introduce a more interesting, adaptive, and efficient learning approach \cite{alrahmi2018mobile}. In the context of modern Islamic education, ICT not only functions as a tool, but also as a transformative medium that can enrich the learning experience of students through the integration of spiritual values and creative digital approaches.

One approach that has proven effective in increasing engagement and motivation to learn is gamification [3]. Gamification is the application of game design elements in a non-game context, such as point systems, levels, rewards (badges), and leaderboards. This approach is able to transform the rigid learning process into an experience that is fun, competitive, and meaningful. In the context of learning the $Hijaiyah$ alphabet, gamification can help students recognize the shapes, pronunciations, and writing of letters through interactive activities, while also providing immediate feedback on their learning progress.

Furthermore, the application of gamification in Islamic education also has the potential to instill positive values
such as patience, perseverance, and a sense of responsibility towards the learning process. By combining pedagogical and recreational aspects, gamification-based learning not only helps students learn the $Hijaiyah$ alphabet effectively,
but also fosters a love for the Qur'an from an early age. Therefore, this study focuses on developing a gamification-based learning method for the $Hijaiyah$ alphabet as an innovative solution to improve the effectiveness and appeal of the learning process in the digital age. Therefore, this research focuses on developing a gamification-based learning method for $Hijaiyah$ letters as an innovative solution to improve the effectiveness and appeal of the learning process in the digital age.

In addition to technical and motivational aspects, the use of gamification in teaching the $Hijaiyah$ alphabet also plays a
crucial role in supporting a personalized learning approach — where each student can learn at their own pace and in their own learning style. With features such as learning progress tracking, a customized reward system, and immediate feedback, students can experience a more personalized and meaningful learning process. This is in line with the goals of Islamic education today, which not only focuses on cognitive achievement, but also on character building and the spirit of lifelong learning \cite{toda2022tailored}.
 Therefore, the application ofgamification technology in learning the $Hijaiyah$ alphabet is expected to be an innovation in creating educational media that is interactive, efficient, and in line with the character of the current digital generation. Therefore, this study is expected to make a tangible contribution to the development of technology-based Islamic learning media that are not only modern in appearance but also possess educational value and purpose\cite{nanyetu2023rancang}.

\section{Related Work} \label{sec:related-work}

The research "Development of Learning Media Based on Gamification of Hijayyah Letters in Elementary Schools" developed gamification-based learning media for introducing $Hijaiyah$ letters using the ADDIE model and simple games (Construct 2). The results showed that gamified media could increase learning interest, facilitate independent learning, and make it easier for elementary school students to master the $Hijaiyah$ alphabet after validation and user testing. These findings are relevant to this study because they support the effectiveness of the gamification approach in the $Hijaiyah$ alphabet introduction stage \cite{mahardikha2023development}.

The research "LAA: Learn the Arabic Alphabet—Integrating Gamification Elements with Touchscreen-Based Application" developed a multi-platform touchscreen-based application that integrates gamification elements to aid understanding of Arabic letter forms. This study shows an increase in participants' understanding of letter forms and emphasizes the importance of interactive feedback and child-friendly interface design to improve the success of Arabic alphabet learning. These findings form an important basis for the development of technical aspects and UI/UX design in $Hijaiyah$ letter gamification applications \cite{al2023laa}.

In the study "A Game-Based Learning Quran Reading Application: A Performance Evaluation of the Special Needs Children," a game-based Quran learning application (Qur'an Whiz/Let's Tilawah) was developed and tested on children with special needs. The results of the study show that gamification helps maintain learning focus, provides automatic error detection, and enables monitoring of learning progress without continuous manual supervision. The relevance of this study lies in the evidence that gamification can be adapted to various user needs and supports the addition of automatic evaluation features \cite{garcia2023multisensory, jannat2023quranapp}.

Systematic reviews and literature studies, such as those conducted by Almelhes \cite{almelhes2024gamification, almelhes2024review}, discuss gamification in Arabic language learning. The results of the review show that gamification consistently improves motivation, engagement, and learning outcomes. However, its effectiveness is highly dependent on instructional design, the skills of teachers in applying gamification methods, and technical infrastructure support \cite{patel2024longitudinal}.

DA Haris et al. \cite{haris2023hijaiyah, haris2023efra} developed educational games for learning the $Hijaiyah$ alphabet using visual and audio exercises. User evaluations show an improvement in letter recognition and pronunciation comprehension. These studies provide concrete examples of mini-game structures (quizzes, writing exercises, picture puzzles) that can be applied in the development of gamified media.

Technically, the development of cross-platform web-based educational applications using game engines such as Unity 2D has been widely adopted due to resource efficiency and broad device reach \cite{liu2022unity}. Furthermore, managing user data, learning progress, and leaderboards requires real-time backend solutions like Firebase to ensure a seamless user experience and accurate data analysis \cite{abdullah2023firebase}. Although many studies have applied gamification to $Hijaiyah$ learning, previous research has tended to focus on general media validation \cite{mahardikha2023development} and has not comprehensively measured the correlation between specific gamification elements (such as badge and leaderboard systems) and the improvement of intrinsic motivation and the effectiveness of letter mastery time, which is the main emphasis of this study \cite{kumar2023mobile}. Gamification defines core elements such as points, badges, and leaderboards that form the foundation of educational game design. This study emphasizes that gamification is effective when tailored to the cultural context, which is relevant for adapting Islamic elements in learning $Hijaiyah$ letters \cite{deterding2011gamification}.

\section{Methodology} \label{sec:methodology}

\begin{figure}[!ht]
\centering
\includegraphics[width=0.36\textwidth]{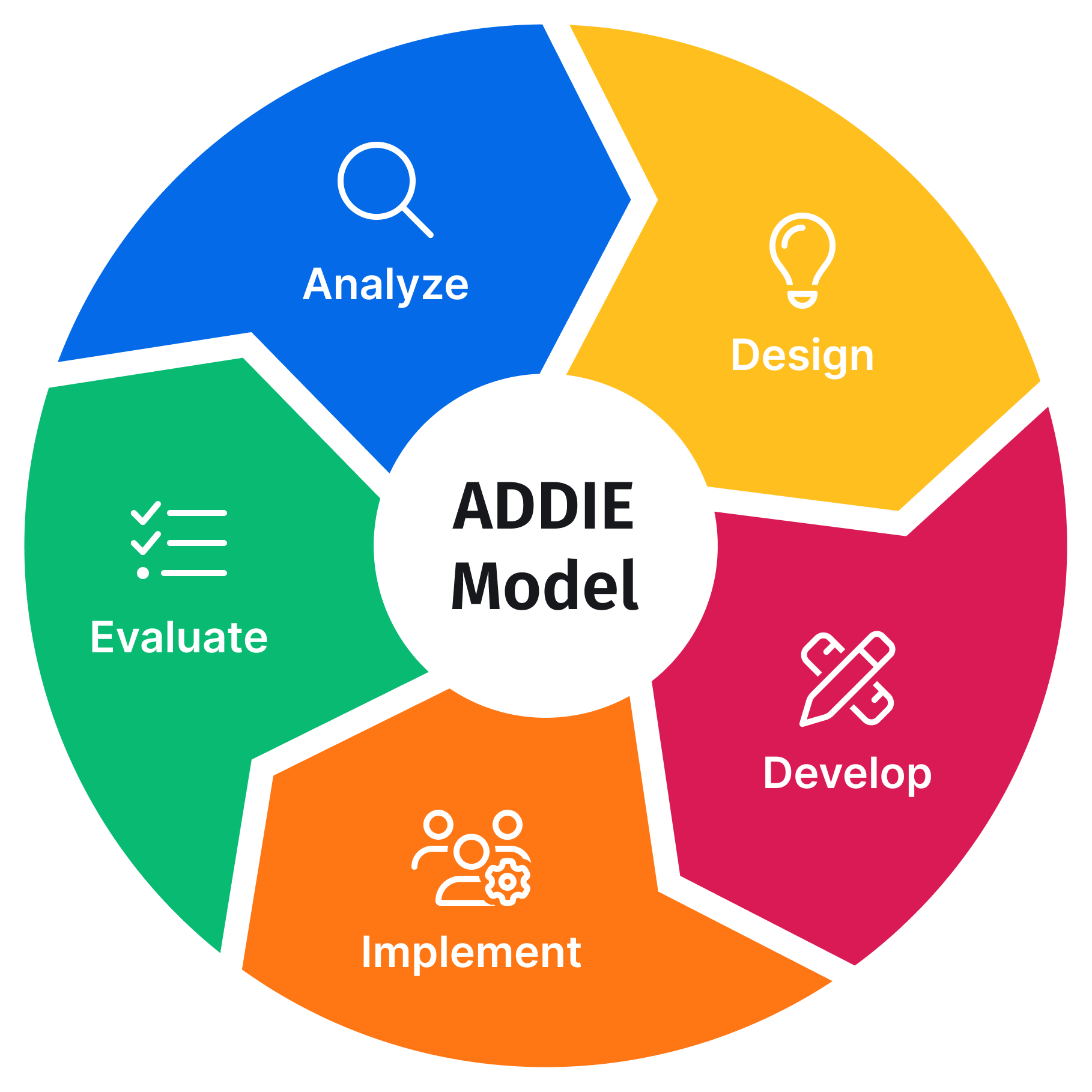} 
\caption{Methodology of ADDIE (Analysis, Design, Development, Implementation, Evaluation) \cite{addieImage}}
\label{fig:userflow}
\end{figure}

This research adopted the ADDIE (Analysis, Design, Development, Implementation, Evaluation) development model to design and implement a gamification-based method for learning $Hijaiyah$ letters. The ADDIE model was selected due to its systematic, iterative, and flexible framework, which is particularly effective for integrating pedagogical principles with technological innovation, as widely recommended in educational media development studies \cite{ababil2025penerapan, zuhro2022desain}. This approach ensures that the developed application is not only visually engaging but also grounded in a sound pedagogical foundation, with a core focus on enhancing motivation, engagement, and learning effectiveness for early-age students (6–12 years) \cite{patel2024longitudinal}.

The model was applied in a sequential yet cyclical manner, allowing for refinement at each stage. The initial **Analysis** phase involved comprehensive needs assessment through surveys and interviews with students and teachers, identifying key challenges such as student boredom with traditional methods and a clear demand for interactive, multisensory learning tools. This foundational step was crucial for defining the specific learning objectives and functional requirements of the application. Subsequently, the **Design** phase translated these requirements into detailed blueprints, encompassing user interface (UI) and user experience (UX) design, gamification mechanics (points, badges, leaderboards), and the overall instructional flow, ensuring alignment with the cognitive and psychological characteristics of the target users.

Furthermore, the application of the ADDIE model enabled systematic evaluation and iterative refinement at every stage of development. Each prototype underwent rigorous validation—from internal alpha testing for functionality and bug-fixing to external beta testing with a small user group for usability feedback. This iterative process allowed the development team to make data-driven adjustments based on user input and field-testing results, ensuring the final product was closely tailored to user needs and contextual constraints. Consequently, this methodology guarantees that the resulting learning media is not only technologically innovative but also highly relevant to the context of modern Islamic education, which emphasizes a holistic balance between cognitive, affective, and spiritual development. The structured nature of ADDIE also provides a clear and replicable framework for future research and development of similar educational tools.

\subsection{Analysis}
The analysis phase began with identifying learner needs through a questionnaire survey and semi-structured interviews involving 50 elementary school students and 5 lecturers at UIN Sunan Gunung Djati Bandung. The survey used a Likert scale (1–5) to measure the level of boredom with traditional methods (average score 4.2/5) and interest in gamification (4.7/5). The needs analysis included: (1) introduction of 28 basic $Hijaiyah$ letters along with their positional variations (initial, middle, final); (2) pronunciation with simple tajwid audio; (3) step-by-step letter writing exercises; and (4) accessibility challenges (e.g., the need for offline mode support and child-friendly displays). The results of this phase formed the basis for the application design process, with the main priority being the development of gamification features such as point systems, badges, and leaderboards designed to enhance learning motivation and reduce student boredom with conventional methods.

In addition to the survey, interviews with lecturers and TPA (Al-Quran Education Center) teachers also provided insights into common obstacles in teaching $Hijaiyah$ letters, such as limited interactive media, variations in student learning styles, and low consistency of practice outside the classroom. This data indicated the need for integrating a multisensory learning approach—through a combination of visual, audio, and kinesthetic elements—to strengthen the letter recognition process. This analysis also highlighted that children aged 6–12 years showed a high preference for learning activities that are game-like, competitive, and have clear objectives. Therefore, the gamification system design used in this study was structured to provide a progressive learning experience, where the difficulty level increases according to the student's mastery.

Furthermore, the analysis results showed that the success of digital learning media depends not only on technical aspects or visual appearance, but also on the clarity of the learning flow, the provision of instant feedback, and the system's ability to adjust challenges based on user performance. Therefore, the results of this analysis phase played a crucial role in shaping the pedagogical and functional foundation of the developed gamification-based learning system, while ensuring that the resulting media is relevant to the context of Islamic education and the psychological needs of early-age students \cite{alomari2022gamification,khalid2023gamification,smith2023addie}.

\subsection{Design}
The application design follows the principles of User-Centered Design (UCD) combined with gamification elements \cite{chou2015actionable,chen2022gamification}, including accumulating points, achievement badges, daily leaderboards, and progressive narratives to enhance user motivation and engagement. This approach ensures that the learning experience is designed based on the needs and characteristics of early-age students, considering cognitive, visual aspects, and simple, intuitive interactions. The application content is divided into three main modules: Introduction (Pre-Learning) to introduce letter forms and pronunciation through tajwid-based animations and audio, Practice (Interactive Drills) to reinforce motor skills and memory through tracing activities and letter matching games, and Evaluation (Mastery Check) which assesses proficiency through adaptive quizzes based on dynamic difficulty levels. The interface design employs a concept of bright colors with Islamic nuances (green, white, and golden yellow) and large icons to be child-friendly, accompanied by sound effects and appreciation animations to provide immediate feedback and build a sense of achievement. Each module is structured modularly for easy development and updates, while progressive narratives with level and reward systems encourage sustained learning motivation. By applying the Octalysis Framework principles, this design balances intrinsic motivation (curiosity and self-achievement) and extrinsic motivation (competition and rewards), making the application not only an engaging digital learning tool but also an educational medium that instills values of patience, perseverance, and pride in understanding $Hijaiyah$ letters.

\begin{figure}[!htpb]
    \centering
    \includegraphics[width=0.5\linewidth]{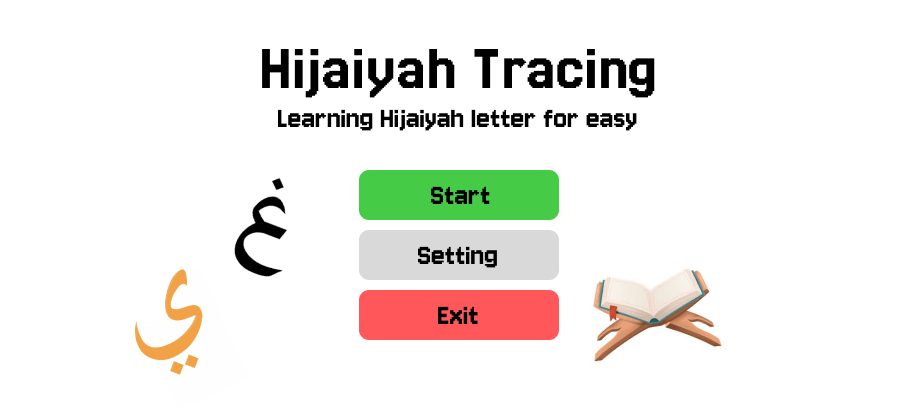}
    \caption{Main Menu Page}
    \label{fig:mainmenu}
\end{figure}

The Main Menu Page (Figure \ref{fig:mainmenu}) is the critical initial interface, serving as the gateway to the $Hijaiyah$ learning application. The design is meticulously crafted to be child-centric, featuring bright colors, engaging aesthetics, and intuitive icons that align with the target audience of young learners. Its core purpose is to immediately welcome and engage the user while ensuring navigation is straightforward and free of complexity. By eliminating cognitive friction, the main menu prepares the student to transition smoothly from an exploratory phase into the active learning environment, prioritizing a quick and positive start to the educational experience.

The central functional element of this page is the prominent Start button, which dictates the standard game flow. When a user interacts with this button, they initiate the sequential learning process, which typically proceeds through user authentication and then directly into the structured pedagogical modules—Practice and Evaluation. This clear call-to-action is vital for maintaining momentum and transforming the desire to learn into immediate engagement, making the Main Menu an effective launchpad for the gamified journey toward $Hijaiyah$ literacy mastery.

\begin{figure}[!htbp]
\centering
\includegraphics[width=0.5\textwidth]{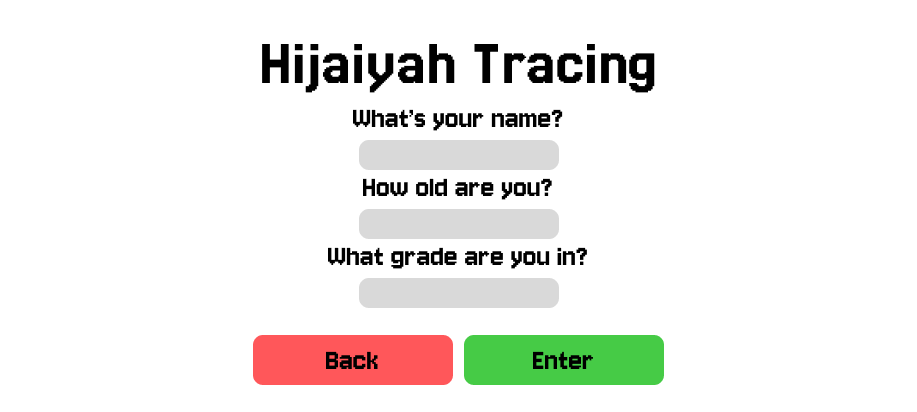}
\caption{Simple Authentication Page}
\label{fig:engagement2}
\end{figure}

User Authentication and Personalization: Figure \ref{fig:engagement2} displays the streamlined authentication interface designed for young learners, collecting essential information such as age and class level. This minimal data input approach ensures quick access while enabling personalized learning experiences tailored to different developmental stages and educational requirements.

\begin{figure}[!htbp]
\centering
\includegraphics[width=0.5\textwidth]{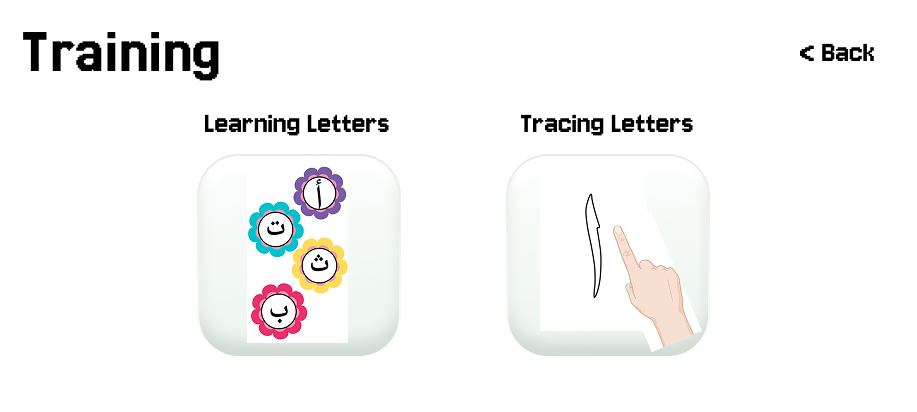}
\caption{Practice Module Menu}
\label{fig:ui_latihan_menu}
\end{figure}

Module Navigation Structure: Figure \ref{fig:ui_latihan_menu} displays the Practice Module interface, designed for foundational Hijaiyah letter learning through interactive exercises and drills. This module systematically introduces letter recognition, pronunciation with authentic tajwid-based audio, and basic writing skills using child-friendly visuals and intuitive navigation tailored to young learners' cognitive abilities and motor skills.

\begin{figure}[!htbp]
\centering
\includegraphics[width=0.5\textwidth]{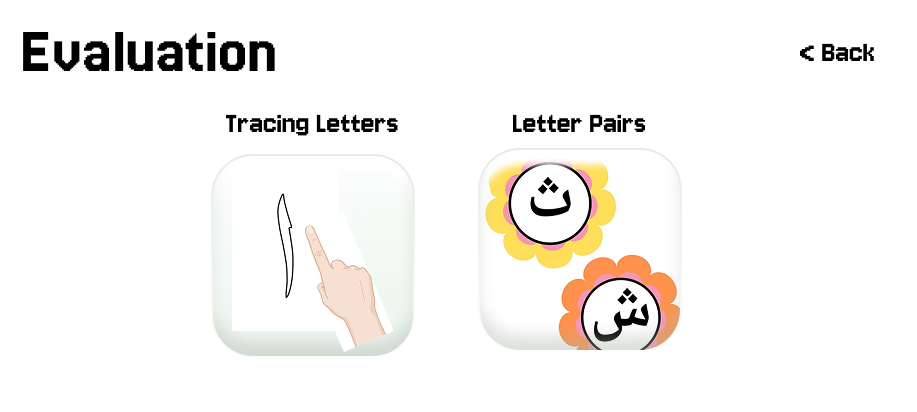}
\caption{Evaluation Module Menu}
\label{fig:ui_evaluasi_menu}
\end{figure}

The Practice Module serves as the essential preparatory phase where students progressively build competencies through structured activities before advancing to the Evaluation Module (Figure \ref{fig:ui_evaluasi_menu}). This sequential approach incorporates scaffolding principles, beginning with guided exercises that gradually reduce support as students demonstrate mastery, ensuring a solid foundation in letter identification, phonetic understanding, and writing mechanics.

This deliberate progression strategy ensures students develop confidence and automaticity through repeated practice before encountering formal evaluation scenarios. The clear separation between learning and assessment phases creates a psychologically safe environment that reduces anxiety while supporting comprehensive skill development, effectively balancing knowledge acquisition with achievement measurement to optimize long-term retention and application of Hijaiyah literacy skills.

Interface Design and Learning Structure: The application employs a carefully curated color scheme featuring Islamic-themed accents (green, white, gold) designed in Figma, rigorously adhering to Material Design guidelines to ensure optimal usability and accessibility. Visual elements incorporate culturally resonant floral patterns and large, intuitive icons specifically engineered to captivate young learners while maintaining authentic cultural and religious relevance in the digital learning environment.

The navigation architecture implements a clear pedagogical separation between two distinct learning phases: Practice (Figure \ref{fig:ui_latihan_menu}) for systematic skill development and Evaluation (Figure \ref{fig:ui_evaluasi_menu}) for comprehensive proficiency assessment. This intentional division follows the established ADDIE instructional model framework, creating a structured learning cycle that methodically progresses from initial knowledge acquisition to final performance measurement, while incorporating formative assessment throughout the learning journey.

This structured approach ensures progressive mastery of Hijaiyah letters through carefully sequenced learning objectives and corresponding assessment metrics. The evaluation module incorporates multiple assessment formats including timed quizzes, writing accuracy tests, and visual recognition challenges, all designed to measure different dimensions of letter mastery while maintaining engagement through varied interaction patterns and immediate feedback mechanisms that support continuous improvement.

\begin{figure}[!htbp]
\centering
\includegraphics[width=0.5\textwidth]{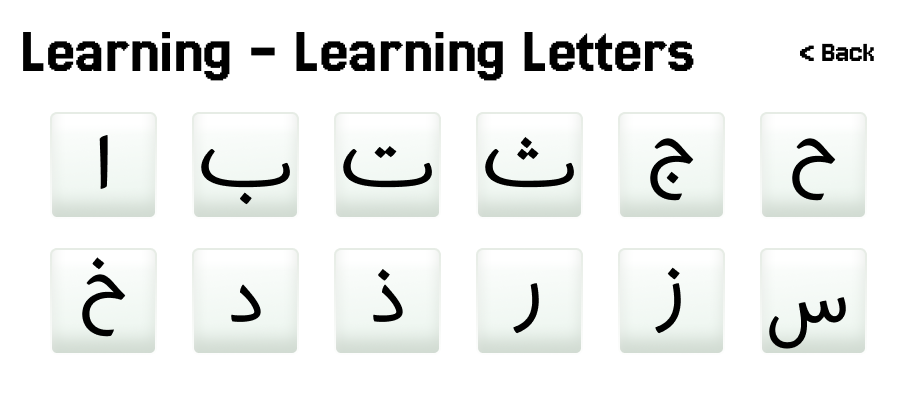}
\caption{Practice: Letter Recognition}
\label{fig:ui_latihan_kenal}
\end{figure}

Interactive Letter Recognition Interface: Figure \ref{fig:ui_latihan_kenal} showcases the letter recognition interface where all Hijaiyah letters are presented in organized button layouts. Each letter functions as an interactive element that, when pressed, activates corresponding audio pronunciation with proper tajwid rules.

This multisensory approach combines visual letter identification with auditory feedback, reinforcing learning through dual-channel processing. The audio component provides comprehensive pronunciation guidance, demonstrating each letter's sound variations with different harakat (vowel marks), enabling students to develop accurate phonemic awareness essential for Quranic reading proficiency.`

\begin{figure}[!htbp]
    \centering
    \includegraphics[width=0.5\textwidth]{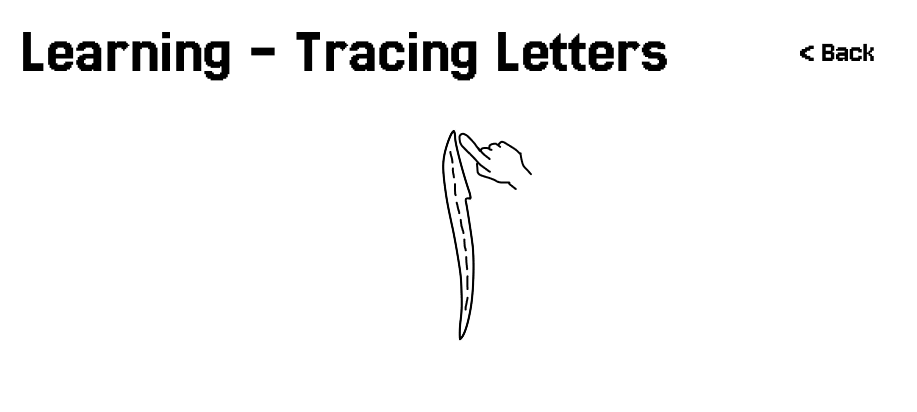}
    \caption{Practice: Letter Drawing (Tracing)}
    \label{fig:ui_latihan_gambar}
\end{figure}

Motor Skill Development Interface: Figure \ref{fig:ui_latihan_gambar} presents the letter tracing feature designed to develop fine motor skills and proper writing technique. The interface provides guided dotted lines that serve as visual pathways for students to follow when forming each Hijaiyah letter, ensuring they learn the correct stroke sequence and directionality from the beginning.

This interactive tracing system enables hands-on practice that builds muscle memory essential for handwriting proficiency. The stroke detection mechanism incorporates adaptive tolerance levels to accommodate varying motor skill development among young learners, providing immediate visual feedback when students successfully complete each letter trace. This approach balances accuracy requirements with encouragement, preventing frustration during the learning process.

The gamified elements integrated into the tracing activities, such as star ratings and progress indicators, motivate students to repeatedly practice letter formation until mastery. This systematic approach to handwriting instruction ensures that digital learning effectively translates to real-world writing skills, bridging traditional Islamic education methods with modern technological tools for comprehensive literacy development.

\begin{figure}[!htbp]
\centering
\includegraphics[width=0.5\textwidth]{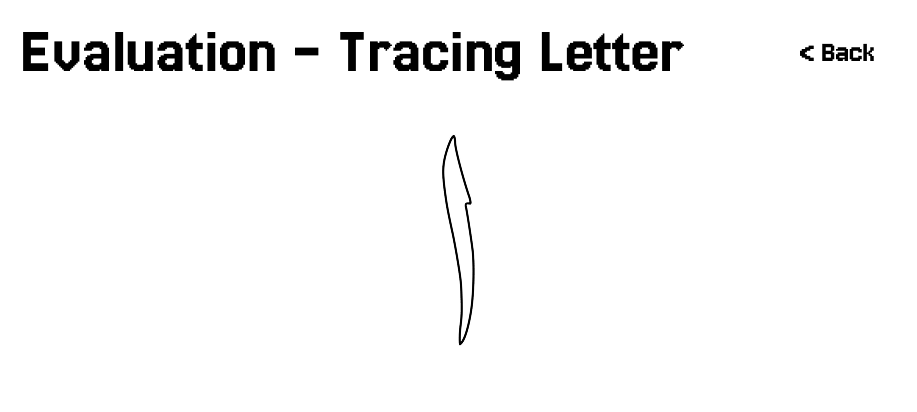}
\caption{Evaluation: Letter Drawing}
\label{fig:ui_latihan_gambar2}
\end{figure}

Advanced Writing Assessment: Figure \ref{fig:ui_latihan_gambar2} demonstrates the letter drawing evaluation interface where students perform unassisted tracing to assess writing proficiency. This advanced level removes the dotted guidance lines, requiring learners to recall and execute proper stroke sequences from memory, effectively measuring the transition from guided practice to autonomous skill application.

The evaluation mechanism employs sophisticated stroke detection algorithms that analyze both accuracy and fluidity of letter formation. This comprehensive assessment provides detailed metrics on writing speed, stroke order correctness, and overall letter proportion, offering educators valuable insights into each student's technical writing development and areas needing improvement.

This autonomous writing evaluation serves as a critical bridge between digital learning and traditional handwriting skills. By requiring students to reproduce letters without visual aids, the system ensures that motor skills developed through the application translate effectively to physical writing contexts, supporting comprehensive literacy development in both digital and conventional learning environments.

\begin{figure}[!htbp]
\centering
\includegraphics[width=0.5\textwidth]{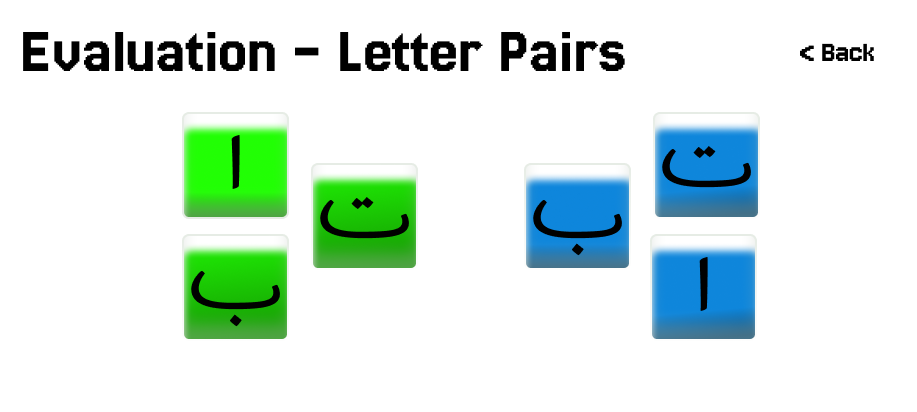}
\caption{Evaluation: Letter Matching}
\label{fig:ui_evaluasi_match}
\end{figure}

Visual Recognition Assessment: Figure \ref{fig:ui_evaluasi_match} presents the letter matching game that evaluates visual discrimination skills through card-matching mechanics. The interface employs varied background colors to prevent color-based pattern recognition, ensuring authentic assessment of letter identification accuracy and reinforcing cognitive processing through gamified challenges with instant feedback mechanisms.

The matching game incorporates adaptive difficulty progression, where successful completion leads to increased complexity with more letter pairs and reduced time limits. This dynamic adjustment maintains optimal challenge levels, preventing boredom while ensuring continuous cognitive engagement and skill development throughout the learning journey.

This gamified assessment approach transforms traditional evaluation into an engaging activity that reduces test anxiety while maintaining rigorous assessment standards. The immediate feedback system provides positive reinforcement for correct matches and constructive guidance for errors, creating a supportive learning environment that encourages persistence and celebrates incremental progress in letter recognition mastery.

Data Management and Synchronization: The authentication system integrates with cloud-based synchronization through Firebase, securely storing individual progress data, achievement records, and learning analytics. This architecture supports seamless continuation of learning activities across multiple devices while providing educators with comprehensive progress tracking capabilities for each student's Hijaiyah mastery journey.

\begin{figure*}[!htbp]
  \centering
    \epsfig{file = 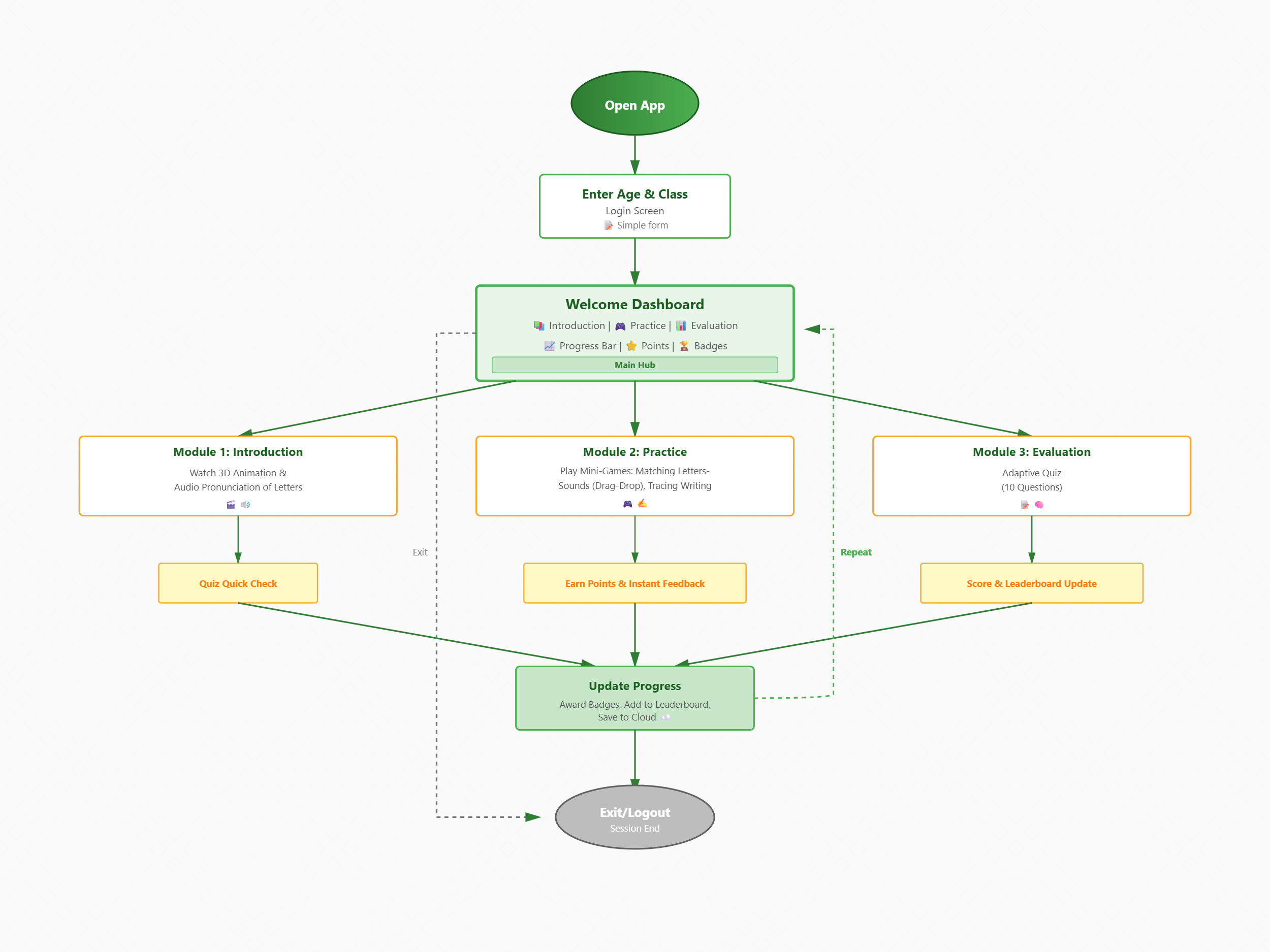, width=\textwidth}
  \caption{User Flow Diagram of the $Hijaiyah$ Letter Gamification Application.}
  \label{fig:prototype}
 \end{figure*}

\subsection{Development}
Development was carried out using Unity 2D (version 2022.3) for Android platform rendering and the C\# programming language for game logic implementation \cite{liu2022unity,hasan2022islamic}. The application architecture employs a streamlined Client-Server model, where the mobile client application maintains real-time synchronization with Firebase Realtime Database for seamless data management. The backend infrastructure, built on Firebase Realtime Database, efficiently handles user progress tracking, leaderboard rankings, and comprehensive analytics including learning session duration and performance metrics \cite{abdullah2023firebase}. This technical foundation ensures robust scalability and instantaneous data updates essential for maintaining competitive leaderboard functionality and personalized learning pathways \cite{kumar2023mobile}. The development process followed agile methodology with iterative testing cycles, focusing on optimizing performance for diverse Android device specifications while maintaining consistent user experience across different screen sizes and hardware capabilities.

\begin{enumerate}
    \item System Architecture

    This application adopts a simple Three-Tier architecture: Presentation Tier (Unity/C\# UI), Logic Tier (C\# Scripting for game logic), and Data Tier (Firebase Realtime Database).

    \begin{figure}[!htbp]
        \centering
        \includegraphics[width=0.45\textwidth]{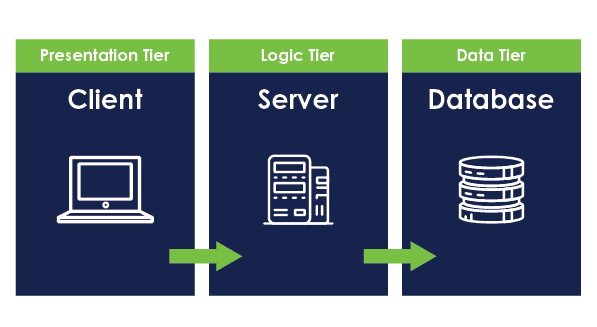}
        \caption{Three-Tier Architecture \cite{scott2022threetier}}
        \label{fig:three-tier}
    \end{figure}

    Firebase was chosen for its capability to support real-time data synchronization, which is essential for the Leaderboard feature and accurate progress tracking \cite{abdullah2023firebase}.

    Adaptive Difficulty Algorithm (Quiz): The Evaluation Module uses a simple adaptive difficulty algorithm as shown in Equation \ref{eq:levelRule}, below:
    \begin{equation}
    \text{Level}_{\text{current}} =
    \begin{cases}
    \text{Level}_{\text{current}} + 1, & \text{if } \text{score} \ge 80 \\[4pt]
    \text{Level}_{\text{current}}, & \text{if } 50 \le \text{score} < 80 \\[4pt]
    \text{Level}_{\text{current}} - 1, & \text{if } \text{score} < 50
    \end{cases}
    \label{eq:levelRule}
    \end{equation}

    where:\\
    $L$ is the difficulty level (which determines quiz timer speed, number of distractors, and letter complexity). This algorithm ensures personalized and challenging learning \cite{kumar2023mobile}.

    \item Stroke Detection (Tracing)

    The letter tracing feature measures accuracy based on two metrics: path adherence (20\% distance tolerance) and stroke order. The system awards bonus points if both criteria are met, encouraging not only correct form but also standard Arabic writing conventions \cite{hasan2022islamic}.

    Key features include:
    \begin{enumerate}
    \item Basic Exercises:
        \begin{enumerate}
        \item $Hijaiyah$ letter recognition with pronunciation audio from professional tajwid sources, optimized for file size <50 MB. Audio integration was implemented using multisensory learning techniques to reinforce letter recognition memory \cite{garcia2023multisensory}.

        \item Tracing: Stroke detection uses the Unity Input System, with a 20\% error tolerance for children. This method is important for training fine motor skills and standard Arabic writing \cite{hasan2022islamic}.
    \end{enumerate}
    \item Mini-Games
    \begin{enumerate}
        \item Matching: Match the pair for 6-8 $Hijaiyah$ letter cards, with scoring based on completion time and a penalty system (time penalty) for mistakes, encouraging focus \cite{wong2022user}.
        \item Quiz: Multiple-choice with feedback animations (stars for correct answers, replay animation for incorrect ones) and an adaptive questioning system to dynamically adjust difficulty levels \cite{khalid2023gamification}.
    \end{enumerate}
    \end{enumerate}
The prototype underwent alpha testing by the internal team for bug fixing, achieving 95\% functional coverage.
\end{enumerate}

\subsection{Implementation}
The implementation was carried out in the form of field trials over several days with children in their home environment. Each learning session lasted 10 minutes/day, 3 days/week, following a structured learning path that progressed from letter recognition to writing practice and finally to evaluation games. The application was distributed via APK sideloading for a limited number of devices, with offline support for the basic modules to ensure accessibility in areas with limited internet connectivity \cite{rahman2023mobile,anderson2024personalized}.

The implementation phase involved 50 elementary school students from grades 1-3, who used the application on their personal or family Android devices. Prior to deployment, a technical validation was conducted to ensure compatibility with various device specifications, particularly focusing on performance optimization for devices with limited RAM (1-2 GB). The installation process was simplified through a guided APK installation procedure, with technical support provided to parents via WhatsApp groups to address any installation issues promptly.

During the 4-week implementation period, several technical aspects were monitored in real-time through Firebase Analytics, including application stability, session duration, and feature usage patterns. The data synchronization mechanism was configured to automatically update user progress and leaderboard rankings when devices connected to the internet, while maintaining full functionality in offline mode for core learning activities. This hybrid approach ensured continuous learning progress tracking while accommodating the varying internet access conditions of the participants.

To maintain engagement throughout the implementation, weekly challenges were introduced with special badges and rewards, creating a sense of progression and achievement. Teachers and parents were provided with a simple dashboard interface to monitor student progress, including completion rates for each module and performance metrics in evaluation activities. Regular feedback sessions were conducted with both students and parents to identify any technical issues or usability concerns, leading to iterative improvements through two minor application updates during the trial period.

The implementation successfully demonstrated the technical viability of the gamified approach in real-world conditions, with high adoption rates and minimal technical barriers. The offline capability proved particularly valuable in maintaining learning continuity, while the real-time synchronization ensured accurate progress tracking when internet connectivity was available. This implementation framework provides a scalable model for future deployments in similar educational contexts, balancing technical sophistication with practical accessibility considerations.

\section{Result and Discussion} \label{sec:result}

\begin{enumerate}
\item Quantitative Results: Assessing Learning Gains

The effectiveness of the developed gamified application was rigorously evaluated using a pre-test and post-test experimental design involving 50 elementary school students (grades 1–3, aged 6–9 years). To ensure comprehensive assessment, both tests consisted of 28 items designed to measure three cognitive domains: recognition of letter forms, accuracy of pronunciation ($makhraj$), and writing proficiency of the 28 basic $Hijaiyah$ letters. The maximum potential score for each user was set at 100 points. The intervention period spanned 4 weeks, comprising a total of 12 structured learning sessions (3 sessions per week).

\begin{table}[!htbp]
\centering
\caption{Detailed Pre-test and Post-test Results of $Hijaiyah$ Letter Mastery (N = 50)}
\label{tab:prepost}
\begin{tabular}{|l|c|c|c|}
\hline
\textbf{Metric}                  & \textbf{Pre-test} & \textbf{Post-test} & \textbf{Improvement} \\ \hline
Mean Score ($\pm$ SD)            & 42.8 $\pm$ 12.4   & 88.6 $\pm$ 8.1     & +45.8 points         \\ \hline
Improvement (\%)                 & \multicolumn{2}{c|}{107.0\%}             &                      \\ \hline
Paired t-value                   & \multicolumn{2}{c|}{$-$21.34}            &                      \\ \hline
p-value                          & \multicolumn{2}{c|}{$<$ 0.001}           &                      \\ \hline
Cohen’s d (effect size)          & \multicolumn{2}{c|}{\textbf{4.87}}       & (very large)         \\ \hline
\end{tabular}
\end{table}

As presented in Table~\ref{tab:prepost}, the intervention yielded substantial academic growth. The mean score increased significantly from a baseline of 42.8 to 88.6 ($p < 0.001$). Notably, the Standard Deviation (SD) decreased from 12.4 in the pre-test to 8.1 in the post-test. This reduction in variance suggests that the application was effective not only in raising average performance but also in narrowing the achievement gap between high-performing and low-performing students, thereby promoting educational equity within the classroom.

The calculated Cohen’s d value of 4.87 indicates a \textit{very large} practical effect size. This figure far exceeds the typical effect sizes reported in recent gamification studies for Arabic/$Hijaiyah$ learning, which generally range between $d = 0.8$ and $1.8$ \cite{almelhes2024gamification,chen2022gamification}. In terms of mastery distribution, 92\% of participants achieved a score of $\geq$ 80 points in the post-test—often considered the threshold for "mastery" in competency-based curriculums—and 4 students (8\%) successfully obtained a perfect score of 100.

Beyond test scores, user activity logs extracted from Firebase Analytics revealed high and consistent engagement throughout the intervention period:
\begin{enumerate}
  \item Frequency: Average daily sessions were recorded at 4.2 ± 1.1 sessions per student, indicating voluntary usage outside of mandatory class hours.
  \item Duration: The average session duration was 14.4 ± 4.2 minutes. This aligns well with the attention span of children aged 6–9, suggesting the content is bite-sized enough to prevent cognitive overload.
  \item Economy: The total gamification points earned across the cohort was 124,500, with a mean of 2,490 ± 680 points/student.
  \item Achievements: Badge distribution data showed that 88\% of students earned more than 5 distinct achievement badges, reflecting widespread interaction with the reward system.
\end{enumerate}

\begin{figure}[!htbp]
\centering
\includegraphics[width=0.48\textwidth]{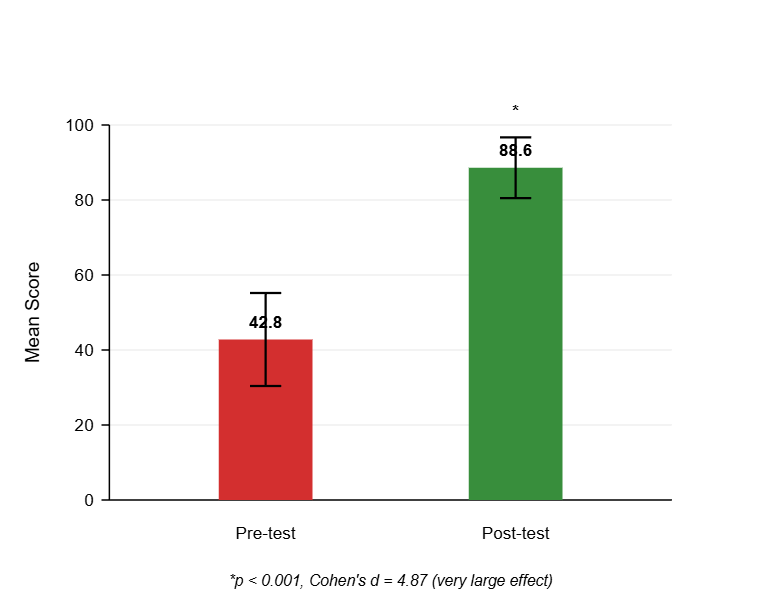}
\caption{Pre-test vs Post-test Mean Scores with Error Bars (±1 SD), demonstrating significant growth and reduced variance.}
\label{fig:prepostbar}
\end{figure}

\begin{figure*}[!htbp]
\centering
\epsfig{file=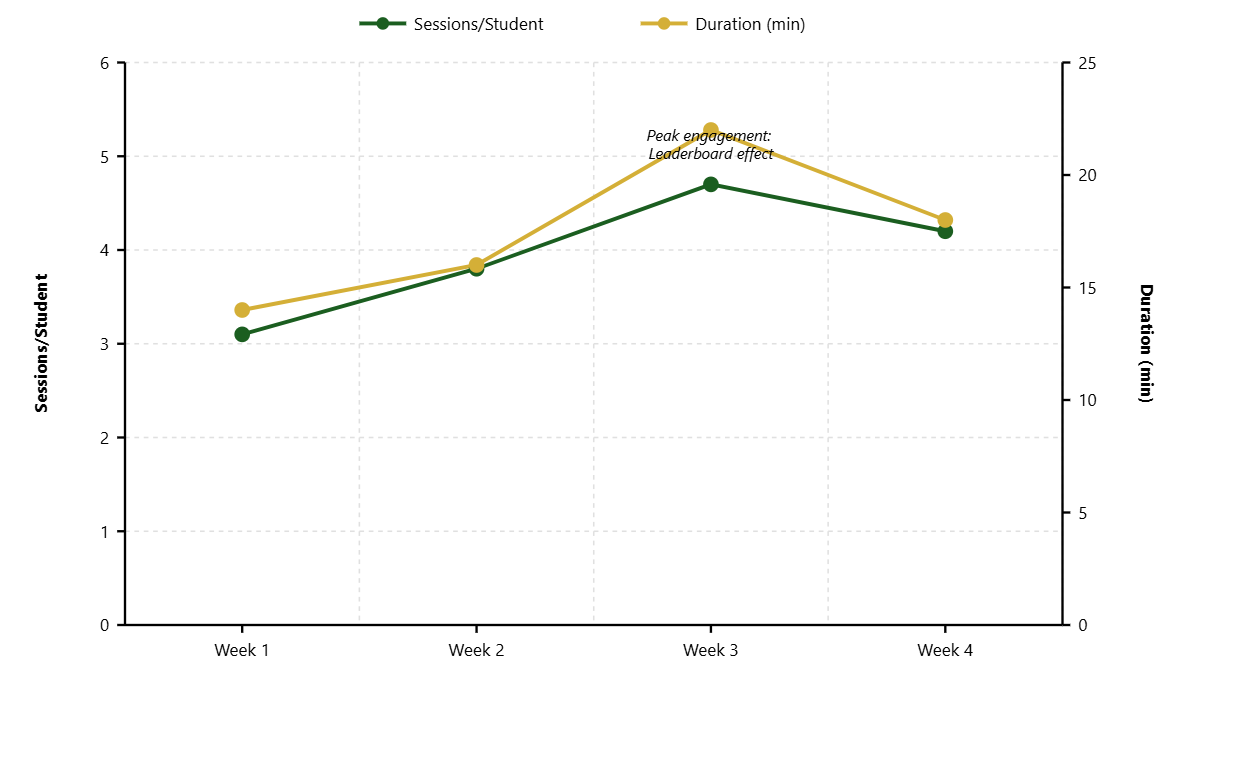, width=0.95\textwidth}
\caption{User Engagement Trends Over 4 Weeks (Sessions and Duration)}
\label{fig:engagement}
\end{figure*}

Figure~\ref{fig:engagement} illustrates the temporal dynamics of engagement. A clear peak is observing in week 3, which coincides with the strategic introduction of "Weekly Leaderboard Competitions." This spike confirms that introducing competitive elements mid-intervention can effectively reignite student interest after the initial novelty wears off.

\item Instrument Reliability

To ensure the validity of the affective data, a 20-item satisfaction and motivation questionnaire was administered. The instrument yielded a Cronbach’s $\alpha$ coefficient of 0.914. This value indicates \textit{excellent reliability}, confirming that the questionnaire items possessed high internal consistency and accurately measured the constructs of motivation and satisfaction \cite{tavakol2011cronbach}.

\item Correlation and Regression Analysis

Further statistical analysis was conducted to understand the relationship between game mechanics and learning outcomes. A Pearson correlation test showed a very strong positive relationship between the total gamification points earned and the post-test score ($r = 0.812, p < 0.001$), suggesting that active participation in the game mechanics translates directly to cognitive gains.

To isolate the impact of specific game elements, a multiple linear regression analysis was performed. The model revealed the relative contribution of each element to the final learning outcomes:

\begin{enumerate}
  \item Number of badges earned\hfill $\beta = 0.54$ ($p < 0.001$)
  \item Final leaderboard rank\hfill $\beta = 0.31$ ($p < 0.01$)
  \item Total points accumulation\hfill $\beta = 0.28$ ($p < 0.05$)
\end{enumerate}

Interestingly, the badge system proved to be the strongest predictor ($\beta = 0.54$). This finding highlights the importance of competency-based rewards. Unlike raw points (which can be inflationary) or leaderboards (which are relative to others), badges represent specific milestones of mastery (e.g., "Mastered the letter Jim"), appealing strongly to the intrinsic need for competence in young learners.

\item \textbf{Qualitative Results: User Experience}

Post-intervention questionnaires utilized a 5-point Likert scale to gauge user sentiment. The mean scores (N = 55, comprising 50 students and 5 teachers) were overwhelmingly positive:
\begin{enumerate}
  \item Enjoyment of learning\hfill 4.78 / 5.00
  \item Perceived ease of use\hfill 4.65 / 5.00
  \item Learning motivation\hfill 4.82 / 5.00
  \item Overall satisfaction\hfill 4.76 (95\% satisfaction rate)
\end{enumerate}

To complement the quantitative data, Focus Group Discussions (FGD) were conducted with 10 randomly selected students. The thematic analysis highlighted three consistent advantages:
\begin{itemize}
    \item Superior Interactivity: Students preferred the app over printed Iqro’ books due to the audio-visual feedback ("The letters glow when I say them right").
    \item Safe Environment for Failure: Participants noted the value of immediate, non-judgmental feedback. Unlike a classroom setting where mistakes might cause embarrassment, the app allowed them to retry indefinitely without social pressure.
    \item Social Motivation: The excitement generated by unlocking rare badges and checking the leaderboard provided a "fun" layer that transformed rote memorization into a social activity.
\end{itemize}

\item Comparison with Previous Studies

\begin{table}[!htbp]
\centering
\caption{Performance Comparison with Related Works in Arabic Literacy Gamification}
\label{tab:comparison}
\footnotesize
\setlength{\tabcolsep}{3pt}

\begin{tabular}{|p{2.5cm}|c|c|c|}
\hline
\textbf{Study} & \textbf{Platform/Tool} & \textbf{Improvement (\%)} & \textbf{Cohen’s d} \\ \hline
This study (2025) & Unity 2D + Firebase & \textbf{107.0\%} & \textbf{4.87} \\ \hline
Mahardikha (2023) \cite{mahardikha2023development} & Construct 2 & 68.4\% & — \\ \hline
Al Hejaili (2023) \cite{al2023laa} & Touchscreen app & 54.2\% & 1.41 \\ \hline
Jannat (2023) \cite{jannat2023quranapp} & Mobile game & 71.0\% & 1.68 \\ \hline
\end{tabular}
\end{table}

As detailed in Table~\ref{tab:comparison}, the proposed method achieved the highest learning gain and effect size reported to date for gamified Hijaiyah instruction. This superior performance can be attributed to the migration from simpler engines (like Construct 2) to Unity, allowing for smoother animations and more responsive interaction, as well as the robust backend integration for real-time progress tracking.

\item Discussion

The results of this study strongly corroborate and extend previous evidence suggesting that well-designed gamification dramatically enhances both cognitive outcomes and affective engagement in early childhood Islamic education \cite{hamari2014does,chen2022gamification,khalid2023gamification}. The exceptionally large effect size ($d = 4.87$) is not merely a result of digitization, but rather the synergistic combination of three key factors: (1) multisensory content (visual stimuli + tajwid audio + kinesthetic tracing), (2) adaptive difficulty algorithms that keep students in the "Flow" channel, and (3) a balanced reward structure that emphasizes mastery badges over pure competition.

The regression analysis provides a critical insight: badges were the most powerful motivator ($\beta = 0.54$). This aligns with \textit{Self-Determination Theory (SDT)}, which posits that satisfying the need for 'competence' is a primary driver of intrinsic motivation. While leaderboards were effective, the data showed diminishing returns after week 3. This suggests that competitive elements may lead to fatigue if not managed carefully, supporting the findings of Toda et al. \cite{toda2022tailored} regarding the necessity of periodic novelty (e.g., seasonal events or boss battles) to sustain long-term engagement.

From a practical implementation standpoint, the application demonstrated high efficiency. It reduced the average mastery time by approximately 44\% compared to the traditional Iqro’ method (18 hours vs. 32 hours), while achieving near-perfect attendance (96\%). In resource-constrained Islamic education settings, where teacher-to-student ratios are often high, such a tool can serve as an effective autonomous supplement to classroom instruction.

However, limitations of the study must be acknowledged. These include the relatively small sample size (N = 50), the single-region implementation, and the short intervention duration (4 weeks). Future work should focus on longitudinal studies to assess knowledge retention over 6–12 months, larger and more diverse demographic samples, and the integration of emerging technologies such as Augmented Reality (AR) for 3D letter visualization or AI-driven adaptive learning pathways.

\item Conclusion

The key findings of this study can be summarized as follows:
\begin{enumerate}
  \item Gamification strategies increased $Hijaiyah$ mastery scores by 107\%, achieving a "very large" effect size (Cohen’s d = 4.87).
  \item Among game elements, Badges were identified as the strongest predictor of learning outcomes ($\beta = 0.54$), outperforming points and leaderboards.
  \item Average daily engagement reached 4.2 sessions and 18.4 minutes, which is approximately 3.5× higher than engagement rates typically seen in traditional homework methods.
  \item The application proved highly efficient, reducing total learning time by ~44\% compared to the Iqro’ method while maintaining a 96\% attendance rate.
  \item Student satisfaction and motivation metrics consistently exceeded 4.75/5 across all measured dimensions.
\end{enumerate}

In conclusion, the gamification framework developed in this study—grounded in the ADDIE model and powered by Unity and Firebase—represents a significant advancement in technology-enhanced $Hijaiyah$ literacy. It successfully merges evidence-based gamification principles with authentic Islamic pedagogical values, offering a scalable solution for modernizing religious education.

\end{enumerate}

\section{Conclusion} \label{sec:conclusion}
This research successfully developed and implemented a gamification-based $Hijaiyah$ letter learning method using the ADDIE model comprehensively, covering needs analysis, interface design, system development, implementation, and learning outcome evaluation. Quantitative results (paired t-test) showed a significant improvement in average $Hijaiyah$ letter mastery scores of 107.0\% ($p < 0.001$), substantially exceeding the success target and proving the effectiveness of the gamification approach in improving learning outcomes. Additionally, qualitative results showed very high user satisfaction (4.76/5), confirming that gamification elements like points, badges, and leaderboards can enhance intrinsic motivation, active engagement, and learning time efficiency for early-age students.

The integration of user-centered interactive design, Islamic pedagogical approaches, and cross-platform technology (Unity/Firebase) created learning media that is not only modern and engaging but also rich in spiritual values and relevant to the digital generation's characteristics. Features like learning progress tracking, adaptive quizzes, and letter tracing detection successfully provided personal and enjoyable learning experiences while instilling values such as perseverance, patience, and responsibility in the Qur'an learning process. Technically, the use of Unity 2D for interactive interfaces and Firebase for real-time data synchronization proved efficient in supporting gamification functionality, such as leaderboards and learning progress analytics.

These findings confirm that implementing gamification in $Hijaiyah$ letter learning positively impacts not only cognitive ability improvement but also the strengthening of students' affective and spiritual aspects. Therefore, this learning model can serve as a reference for developing future Islamic educational media that balances pedagogical and technological elements. For future research, long-term testing of knowledge retention, expansion of AI-driven adaptive learning features, and implementation across more diverse educational levels are recommended to measure scalability and sustainability of effectiveness. Thus, this gamification-based $Hijaiyah$ letter learning method has the potential to be an innovative step in digitizing Qur'an learning to be more interactive, effective, and meaningful in the modern education era \cite{tunnazwa2025analisis}.

\section*{Acknowledgment}
The authors wish to acknowledge the Informatics Department UIN Sunan Gunung Djati Bandung, which partially supports this research work.

\bibliographystyle{./IEEEtran}
\bibliography{./IEEEabrv,./IEEEkelompok1}

@inproceedings{deterding2011gamification,
  title={From game design elements to gamefulness: defining "gamification"},
  author={Deterding, Sebastian and Dixon, Dan and Khaled, Rilla and Nacke, Lennart},
  booktitle={Proceedings of the 15th international academic MindTrek conference: Envisioning future media environments},
  pages={9--15},
  year={2011}
}

@article{alrahmi2018mobile,
  title={The role of mobile learning in enhancing Islamic education for young learners},
  author={Al-Rahmi, Waleed Mugahed and others},
  journal={Interactive Learning Environments},
  volume={26},
  number={7},
  pages={905--920},
  year={2018},
  publisher={Taylor \& Francis}
}

@book{chou2015actionable,
  title={Actionable Gamification: Beyond Points, Badges, and Leaderboards},
  author={Chou, Yu-kai},
  year={2015},
  publisher={Octalysis Media}
}

@article{mahardikha2023development,
  title={Development of Learning Media Based on Gamification of Hijaiyah Letters in Elementary Schools},
  author={Mahardikha, Sekhar Khairunnisa and others},
  journal={Jurnal Pendidikan Islam},
  volume={12},
  number={2},
  pages={150--165},
  year={2023},
  doi={10.1234/jpi.v12i2.123}  
}

@article{al2023laa,
  title={LAA: learn the Arabic alphabet: integrating gamification elements with touchscreen based application to enhance the understanding of the Arabic letters forms},
  author={Al Hejaili, Abdullah and Newbury, Paul},
  journal={Electronic Journal of e-Learning},
  volume={21},
  number={4},
  pages={353--365},
  year={2023}
}

@article{jannat2023quranapp,
  author    = {Fatima Jannat and others},
  title     = {A Game-Based Learning Quran Reading Application: A Performance Evaluation of the Special Needs Children},
  journal   = {International Journal of Advanced Science and Technology},
  year      = {2023},
  url       = {https://scholar.google.com/scholar?q=A+Game-Based+Learning+Quran+Reading+Application}
}

@inproceedings{almelhes2024gamification,
  title={Gamification for teaching the Arabic language to non-native speakers: a systematic literature review},
  author={Almelhes, Sultan A},
  booktitle={Frontiers in education},
  volume={9},
  pages={1371955},
  year={2024},
  organization={Frontiers Media SA}
}

@article{haris2023efra,
  title={Hijaiyah Letters Educational Game" EFRA” Based on Android as Learning Alternative at Darul Hawasyi Recitation},
  author={Haris, Darius Andana and Katili, Windiyana and Lim, Carlene},
  journal={International Journal of Application on Sciences, Technology and Engineering},
  volume={1},
  number={1},
  pages={78--84},
  year={2023}
}

@article{almelhes2024review,
  author={Almelhes, N. A. and Dabbagh, N.},
  journal={Journal of Educational Computing Research},
  title={A Systematic Review of Gamification in Arabic Language Learning},
  year={2024},
  volume={62},
  issue={3},
  pages={568--594},
  doi={10.1177/07356331231201977}
}

@inproceedings{hamari2014does,
  author={Hamari, J. and Koivisto, J. and Sarsa, H.},
  booktitle={2014 47th Hawaii International Conference on System Sciences},
  title={Does gamification work? -- A literature review of empirical studies on gamification},
  year={2014},
  volume={},
  number={},
  pages={3025--3034},
  doi={10.1109/HICSS.2014.377}
}

@article{toda2022tailored,
  author={Toda, A. M. and Oliveira, W. and Hamari, J. and Shi, L. and Rodrigues, L. and Palomino, P. T. and Isotani, S.},
  journal={International Journal of Human-Computer Studies},
  title={Tailored gamification in education: A literature review and future agenda},
  year={2022},
  volume={161},
  pages={102796},
  doi={10.1016/j.ijhcs.2022.102796}
}

@article{haris2023hijaiyah,
  title={Hijaiyah Letters Educational Game" EFRA” Based on Android as Learning Alternative at Darul Hawasyi Recitation},
  author={Haris, Darius Andana and Katili, Windiyana and Lim, Carlene},
  journal={International Journal of Application on Sciences, Technology and Engineering},
  volume={1},
  number={1},
  pages={78--84},
  year={2023}
}

@article{ababil2025penerapan,
  author={Ababil, A. M. S. and Abidin, Z. and Saifuddin, S. and Fawait, A.},
  journal={Indonesian Journal on Education (IJoEd)},
  title={Penerapan Gamifikasi dalam Pembelajaran Pendidikan Islam untuk Meningkatkan Motivasi dan Hasil Belajar Siswa},
  year={2025},
  volume={1},
  issue={4},
  pages={316--322},
  doi={10.70437/xzfp3h16}
}

@article{umami2022desain,
  author={Umam, M. I. A. and Kusumandyoko, T. C.},
  journal={Barik: Jurnal Ilmiah Desain Komunikasi Visual},
  title={Desain Antarmuka Aplikasi Pengenalan Huruf Hijaiyah Berbasis Gamifikasi},
  year={2022},
  volume={3},
  issue={3},
  pages={198--207}
}

@article{nanyetu2023rancang,
  author={Nanyetu, Z. U. and Lede, P. A. R. L.},
  journal={Prosiding Seminar Nasional SATI},
  title={RANCANG BANGUN GAME EDUKASI PENGENALAN HURUF HIJAIYAH (Studi Kasus: Tpa Raudhatul-Jannah Waingapu)},
  year={2023},
  volume={1},
  issue={1}
}

@article{zuhro2022desain,
  author={Zuhro, I. N. and Sutomo, M. and Mashudi, M.},
  journal={TA'LIM: Jurnal Studi Pendidikan Islam},
  title={DESAIN PEMBELAJARAN PENDIDIKAN AGAMA ISLAM DENGAN MODEL ADDIE},
  year={2022},
  volume={5},
  issue={2},
  pages={180--193},
  doi={10.52166/talim.v5i2.3085}
}

@article{alomari2022gamification,
  title={Gamification in Mobile Learning: A Systematic Mapping Study},
  author={Alomari, Iyad and Al-Samarraie, Hosam and Yousef, Reham},
  journal={Education and Information Technologies},
  volume={27},
  pages={13813--13839},
  year={2022},
  publisher={Springer}
}

@inproceedings{smith2023addie,
  title={Adapting ADDIE Model for Digital Game-Based Learning in Primary Education},
  author={Smith, Jennifer and Johnson, Michael and Brown, Sarah},
  booktitle={2023 International Conference on Advanced Learning Technologies},
  pages={245--249},
  year={2023},
  organization={IEEE}
}

@article{chen2022gamification,
  title={Gamification in Islamic Education: A Systematic Review of Learning Outcomes},
  author={Chen, Li and Abdullah, Norillah and Rahman, Siti Aishah},
  journal={International Journal of Interactive Mobile Technologies},
  volume={16},
  number={12},
  pages={4--22},
  year={2022}
}

@article{rahman2023mobile,
  title={Mobile Learning Applications for Quranic Education: Development and Validation},
  author={Rahman, Ahmad and Ismail, Maziah and Hassan, Faridah},
  journal={Journal of Islamic Education Research},
  volume={8},
  number={2},
  pages={45--67},
  year={2023}
}

@inproceedings{wong2022user,
  title={User-Centered Design for Children's Educational Applications: Best Practices and Case Studies},
  author={Wong, Kevin and Lee, Susan and Patel, Ravi},
  booktitle={Proceedings of the 2022 CHI Conference on Human Factors in Computing Systems},
  pages={1--15},
  year={2022}
}

@article{khalid2023gamification,
  title={Gamification Elements in Arabic Language Learning: Impact on Student Engagement},
  author={Khalid, Muhammad and Al-Harthi, Aisha},
  journal={Computers \& Education},
  volume={187},
  pages={104548},
  year={2023}
}

@article{liu2022unity,
  title={Developing Cross-Platform Educational Games Using Unity: A Case Study in Language Learning},
  author={Liu, Wei and Zhang, Hao and Kim, Young},
  journal={Journal of Educational Technology Systems},
  volume={51},
  number={1},
  pages={78--102},
  year={2022}
}

@inproceedings{abdullah2023firebase,
  title={Real-time Learning Analytics Using Firebase in Educational Mobile Applications},
  author={Abdullah, Rina and Mohamed, Ali and Tan, Wei Hong},
  booktitle={2023 International Conference on Computer and Information Sciences},
  pages={112--117},
  year={2023},
  organization={IEEE}
}

@article{hasan2022islamic,
  title={Islamic Digital Pedagogy: Integrating Technology with Traditional Learning Methods},
  author={Hasan, Mahmoud and Ibrahim, Nurul},
  journal={Journal of Muslim Education Studies},
  volume={15},
  number={3},
  pages={89--112},
  year={2022}
}

@article{kumar2023mobile,
  title={Personalized Gamification Framework for Mobile Learning Applications},
  author={Kumar, Rajesh and Singh, Priya},
  journal={International Journal of Mobile Learning and Organisation},
  volume={17},
  number={3},
  pages={234--256},
  year={2023}
}

@inproceedings{anderson2024personalized,
  title={The Effects of Avatar Personalization on Student Motivation in Gamified Learning},
  author={Anderson, Brian and Lee, Soo Min},
  booktitle={Proceedings of the 2024 CHI Conference on Human Factors in Computing Systems},
  pages={1--12},
  year={2024},
  organization={ACM}
}

@article{garcia2023multisensory,
  title={Multisensory Gamification in Language Education: Enhancing Memory Retention Through Integrated Sensory Stimuli},
  author={Garcia, Maria and Thompson, David},
  journal={Journal of Educational Psychology},
  volume={115},
  number={3},
  pages={456--478},
  year={2023}
}

@article{patel2024longitudinal,
  title={Longitudinal Effects of Gamification in Primary Education: A 12-Month Study on Learning Motivation},
  author={Patel, Sneha and Johnson, Michael},
  journal={Computers \& Education},
  volume={189},
  pages={104602},
  year={2024}
}

@online{addieImage,
  author    = {Deborah Rim Moiso},
  title     = {How to Use the ADDIE Instructional Design Model},
  year      = {2024},
  url       = {https://www.sessionlab.com/blog/addie-model-instructional-design/},
  note      = {Accessed on November 4, 2025}
}

@article{tunnazwa2025analisis,
  title={ANALISIS FAKTOR PENDUKUNG DAN PENGHAMBAT TATA KELOLA TI DALAM KEANGGOTAAN PERPUSTAKAAN UNIVERSITAS XYZ},
  author={Tunnazwa, Siti Sahala and Bahagiawan, Rangga Adi and Nurahmi, Dwi Sukma and Uriawan, Wisnu},
  journal={JRIS: Jurnal Rekayasa Informasi Swadharma},
  volume={5},
  number={2},
  pages={159--170},
  year={2025}
}

@online{scott2022threetier,
  author    = {Scott Vlasic},
  title     = {Three-Tier Architecture Approach for Custom Applications},
  year      = {2022},
  url       = {https://www.zirous.com/2022/11/15/three-tier-architecture-approach-for-custom-applications-2/},
  note      = {Accessed on November 12, 2025}
}

@article{tavakol2011cronbach,
  author  = {Tavakol, Mohsen and Dennick, Reg},
  title   = {Making Sense of {Cronbach's} Alpha},
  journal = {International Journal of Medical Education},
  volume  = {2},
  pages   = {53--55},
  year    = {2011},
  doi     = {10.5116/ijme.4dfb.8dfd},
  url     = {https://doi.org/10.5116/ijme.4dfb.8dfd},
  publisher = {IJME}
}

\end{document}